\documentclass[12pt]{article}

\newcommand{\be}{\begin{equation}}
\newcommand{\ee}{\end{equation}}
\newcommand{\bea}{\begin{eqnarray}}
\newcommand{\eea}{\end{eqnarray}}
\newcommand{\nn}{\nonumber \\}
\newcommand{\p}[1]{(\ref{#1})}
\newcommand{\ba}{\begin{array}}
\newcommand{\ea}{\end{array}}
\newcommand{\vs}[1]{\vspace{#1 mm}}

\renewcommand{\a}{\alpha}
\renewcommand{\b}{\beta}
\renewcommand{\c}{\gamma}

\newcommand{\e}{\epsilon}

\def\bbox{{\,\lower0.9pt\vbox{\hrule \hbox{\vrule height 0.2 cm
\hskip 0.2 cm \vrule height 0.2 cm}\hrule}\,}}
\newcommand{\dsl}{\pa \kern-0.5em /}

\newcommand{\pa}{\partial}

\def\a{\alpha}\def\b{\beta}
\def\s{\sigma}

\begin{document}

\topmargin 0pt
\oddsidemargin 5mm

\renewcommand{\thefootnote}{\fnsymbol{footnote}}
\begin{titlepage}

\setcounter{page}{0}

\rightline{\small hep-th/9906162 \hfill QMW-PH-99-08}

\vs{15}
\begin{center}
{\Large Membranes on Fivebranes}
\vs{10}

{\large
Jerome P. Gauntlett } \\
\vs{5}
{\em Department of Physics\\
       Queen Mary and Westfield College\\
       University of London\\
       Mile End Road\\
       London E1 4NS, UK\\
       J.P.Gauntlett@qmw.ac.uk}\\[2em]
\end{center}
\vs{7}
\centerline{{\bf Abstract}}
By analysing supersymmetry transformations we derive 
new BPS equations for the D=11 fivebrane propagating in 
flat space that involve the world-volume three-form. 
The equations generalise those of 2,3,4 and 5 dimensional 
special Lagrangian submanifolds and are relevant for 
describing membranes ending on these submanifolds.

\end{titlepage}
\newpage
\renewcommand{\thefootnote}{\arabic{footnote}}
\setcounter{footnote}{0}

\section{Introduction}
The world-volume theories of branes in string theory and M-theory
are an interesting setting for
studying BPS equations. One notable feature is that 
the solutions incorporate a spacetime interpretation \cite{cm,gib,hlw}
which arises from the fact that the scalar fields 
describe the embedding of the brane in a target space.
The world-volume theory of the D=11 fivebrane, for example,
has a BPS self-dual string soliton that preserves 1/2 of the world-volume
supersymmetry, or equivalently, 1/4 of the spacetime supersymmetry \cite{hlw}. 
It is charged with respect to the world-volume self-dual three-form 
and one scalar field is excited that is harmonic in four variables. 
The simplest solution with a single centre can then be visualised as 
a semi-infinite membrane ending on the fivebrane.
This configuration can be represented as the array
\be\label{first}
\matrix{
M5:&1&2&3&4&5& & & & & \cr
M2:&1& & & &&6&& & & \cr},
\ee
with the self-dual string lying in the 1 direction and the 6 direction
corresponding to the excited scalar field.

Here we will derive the BPS equations corresponding to adding more
membranes to this array. Specifically, we will analyse
arrays with 2,3,4 and 5 orthogonally intersecting membranes,
lying in the $\{2,7\}$, $\{3,8\}$, $\{4,9\}$ and $\{5,10\}$
directions, which 
preserve $1/8$, $1/16$, $1/32$ and $1/32$ of the spacetime supersymmetry,
respectively. The BPS equations are obtained by analysing the supersymmetry
transformations of the fivebrane world-volume theory in a flat target space.
As in  \cite{glw,glwtwo}, 
the array of orthogonally intersecting branes provides a guide 
in constructing the BPS equations. More specifically,
each membrane in the array of intersecting branes suggests which
projections to impose on the supersymmetry spinor parameters and
in addition which scalar fields should be active. This information
then leads to the BPS equations.

We will argue that the equations
are appropriate for describing a membrane ending on a 
fivebrane world-volume, some of which is a special Lagrangian submanifold.
To see this, we first note that a consequence of the projections imposed on
the spinor parameters arising from the arrays
of membranes orthogonally intersecting a single fivebrane, is that 
one can add certain fivebranes to the array ``for free'', 
i.e., without imposing further conditions on the spinor parameters.
{}Following \cite{bbs,bbmooy}, it was shown in \cite{glw} 
(see also \cite{gp,afs}) that an
array of intersecting
fivebranes only corresponds to BPS equations on the world-volume 
of a single fivebrane that imply that the fivebrane world-volume 
(or part of it) is a calibrated submanifold \cite{hl}.
By adding extra membranes to the array (\ref{first}) we find that
the fivebranes we can add for free 
correspond to the special Lagrangian calibrations.
More precisely, adding 2,3,4 and 5 membranes corresponds to the projections
for fivebranes related to 2,3,4 and 5 dimensional special Lagrangian manifolds,
respectively. The full BPS equations will involve the world-volume
three-form and can be interpreted as membranes ending on such fivebrane 
world-volumes\footnote{Note that BPS equations for other configurations
of intersecting fivebranes and membranes were derived in \cite{glwtwo}.}.
It should be emphasised that since the BPS equations incorporate back reaction,
the fivebrane world-volume will in general no longer be special Lagrangian.
Both the geometry and the topology could be different. 

Let us illustrate this interpretation
further for the simplest case. As we will later see explicitly, 
adding a membrane in the $\{2,7\}$ directions to the 
array (\ref{first}) gives rise to
an extra projection on the supersymmetry parameters which implies
that we can add an 
anti-fivebrane in the $\{3,4,5,6,7\}$ directions for free to obtain the array:
\be\label{second}
\matrix{
M5:&1&2&3&4&5& & & & & \cr
M2:&1& & & & &6& & & & \cr
M2:& &2& & & & &7& & & \cr
\overline{M5}:& & &3&4&5&6&7& & & \cr}.
\ee
The two intersecting fivebranes alone correspond to a two-dimensional
special Lagrangian submanifold, or equivalently, a complex curve in  
the $\{1,2,6,7\}$ directions. 
This is the M-theory setup that has been used to analyse $N=2$ 
superYang-Mills (SYM) theory \cite{wit}. 
By including the membranes, the corresponding BPS equations  
will have solutions that can be interpreted as a membrane ending 
on this curve. 
In the Yang-Mills setting these configurations will correspond to 
monopoles and dyons. Such membranes were first considered in \cite{fs,hy,m}
without including the back-reaction on the fivebrane. 
The BPS equations that we discuss here
which do include the back-reaction were first presented in 
\cite{glwtwo,lwmon} and the application to dyons was analysed in \cite{lwmon}.
In \cite{glwtwo,lwmon} the BPS equations were derived and checked 
in a certain approximation in the ``covariant formalism'' of the fivebrane
\cite{hsw}.
Here we shall derive the exact BPS equations
in the Hamiltonian formalism \cite{bergtown}. 
A consequence of this is that we will be able to 
present the exact energy functional for these configurations that was
presented in approximate form in \cite{blw}.

After discussing this case we will 
consider arrays with three intersecting membranes.
We will see that the projections for the membranes imply
the projections corresponding to a three dimensional special Lagrangian
submanifold. Such submanifolds are relevant for describing
N=2 SYM in d=3 \cite{klm} and the equations for the
membranes correspond to solitonic excitations in these theories. 
The BPS equations that we derive provide a 
starting point to analyse such states.

The projections for four intersecting membranes automatically
imply those for five so we will consider these cases together.
The resulting BPS equations are appropriate for studying 
membranes ending on four or five dimensional special Lagrangian 
submanifolds. The BPS equations for the five membrane case are in fact
the most general in the sense that by setting certain scalar fields to
zero we can recover the BPS equations for all previous cases.
In all cases the equations are rather complicated and (almost) no
attempt will be made for finding solutions here.

\section{Two Intersecting Membranes}

To derive the BPS equations we will use 
the Hamiltonian formalism of the fivebrane theory \cite{bergtown} 
which is derived from the ``Lagrangian'' formalism of \cite{pst}.
The bosonic world volume fields consist of scalars $X^\mu$, $\mu=0,\dots,10$,
and a closed three form field-strength 
$H$ that satisfies a self-duality condition.
In the static gauge we have $X^0=\sigma^0$, $X^a=\sigma^a$, $a,b,\dots
=1,\dots, 5$,
where $(\sigma^0,\sigma^a)$ are coordinates on the fivebrane world-volume.
We will only consider static, bosonic configurations\footnote{In the
Hamiltonian formalism we only need to specify the spatial components of
$H$.} $X^T(\sigma^a)$,
$T=6,\dots, 10$, $H_{abc}(\sigma^a)$, and we shall take a flat target space.
The spatial components of the world-volume metric, $g_{ab}$, 
are then given by
\be
g_{ab}=\delta_{ab}+\pa_a X^T \pa_b X^T.
\ee 

It was shown in \cite{glwtwo} that bosonic configurations
that preserve supersymmetry must satisfy
\be\label{bound}
{\sqrt{{\rm det}(g+\tilde H)_{ab}}}=
\epsilon^\dag
\gamma^0\left[{1\over 5!}\Gamma_{a_1\dots a_5} \epsilon^{a_1\dots a_5}
 - {{\sqrt {g}}\over 2 }
\Gamma_{ab} \tilde H^{ab}
+\Gamma_a t^a
\right]\e\ ,
\ee
where
\bea
\tilde H^{ab} &=& {1\over 3!}
{1\over \sqrt{g}}\varepsilon^{abc_1c_2c_3}
H_{c_1c_2c_3} \ ,
\nn
t_f &=& {1\over 4!} \varepsilon^{abcde} H_{abc}H_{def}\ ,
\eea
and $\e$ is a 32 component Majorana spinor satisfying $\e^\dagger \e=1$.
We use the convention that $\varepsilon^{12345}=1$.
The gamma-matrices $\Gamma_a$ are the flat spacetime matrices, $\gamma_a$, 
pulled back to the world-volume:
\be 
\Gamma_a=\gamma_a + \pa_aX^T\gamma_T\ .
\ee
It is useful to introduce the density 
\be
\tilde {\cal H}^{ab}={\sqrt g} \tilde H^{ab}
\ee
so that the closure of the three-form $H$ is then 
$\pa_a \tilde {\cal H}^{ab}=0$.

We now turn to the arrays of
two membranes intersecting a fivebrane according to the pattern
(\ref{second}).
To preserve supersymmetry we must impose the following
projections on the supersymmetry parameters:
\bea\label{gamproj}
\gamma^{012345}\epsilon &=&\epsilon\nn
\gamma^{016}\epsilon&=&\epsilon\nn
\gamma^{027}\epsilon&=&\epsilon
\eea
and hence the configuration preserves 1/8 of the supersymmetry.
These projections imply that 
$\gamma^{034567}\epsilon =-\epsilon$ which means that we can indeed include  
the final fivebrane in (\ref{second}) for free, as claimed earlier. 
To realise this configuration of intersecting branes
as a supersymmetric configuration of the first fivebrane
we set all scalar fields to zero except for 
$X^6\equiv Y^1$ and $X^7\equiv Y^2$, which will be functions
of $\sigma^a=(\s^i,\s^\a)$, where, in this section,
$i,j,\dots =1,2$ and $\a,\b,\dots=3,4,5$.
By substituting this ansatz into (\ref{bound}) and imposing the
projections (\ref{gamproj}) we will be able to derive the BPS
equations.

It will be convenient to introduce the notation 
$(\rho^1,\rho^2)$=$(\gamma^6,\gamma^7)$, so that
projections on the spinor imply that
\bea
\gamma_1\rho_1\e&=&\gamma_2\rho_2\e=\gamma_{12345}\e\nn
\gamma_1\rho_2\e&=&-\gamma_2\rho_1\e\ .
\eea
We then find that
\bea\label{eqone}
\Gamma_{12345}\e&=&[(1-\det\pa_i Y^j)
\gamma_{12345} +\pa_i Y^i
+\pa_iY^j\e_{ij}\gamma_{12} + \pa_\a Y^i
\gamma_{\a i}\nn &+&
(\pa_\a Y^j\pa_j Y^i-\pa_\a Y^i\pa_j Y^j)\gamma_{\a}\rho_i
+\e_{\a\b\c}\pa_\b Y^1 \pa_\c Y^2\gamma_{\a}]\e\ .
\eea
Similarly
\bea\label{eqtwo}
-{1\over 2}{\tilde{\cal H}}^{ab}\Gamma_{ab}\e&=&
[({\tilde{\cal H}}^{12}\e_{ij} \pa_i Y^j + 
{\tilde{\cal H}}^{\a i}\pa_\a Y^i)\gamma_{12345}
+(\pa_\a Y^k{\tilde{\cal H}}^{\a i}\pa_i Y^l \e_{kl}
-{\tilde{\cal H}}^{12}(1-\det \pa_i Y^j))\gamma_{12}\nn
&-&{\tilde{\cal H}}^{\a i}\gamma_{\a i}
-{\tilde{\cal H}}^{\a j}\pa_j Y^i \gamma_\a\rho_i
+({\tilde{\cal H}}^{12}\pa_i Y^i - 
{\tilde{\cal H}}^{\a [i}\pa_\a Y^{j]}\e_{ij})
\gamma_2\rho_1 -{\tilde{\cal H}}^{\a \b}\Gamma_{\a\b}]\e
\eea
and
\be\label{eqthree}
t^a\Gamma_a\e=t^a\gamma_a\e +t^a\pa_a Y^i\rho_i\e\ .
\ee

To ensure the equality of the two sides of (\ref{bound}) the sum of
(\ref{eqone})-(\ref{eqthree}) must vanish except for terms proportional
to $\gamma_{12345}\e$. We conclude that
\bea\label{bpsone}
{\tilde{\cal H}}^{\a \b}&=&0\nn
{\tilde{\cal H}}^{\a i} &=&\pa_\a Y^i\nn
{\tilde{\cal H}}^{12}&=&{ \e_{ij}\pa_i Y^j +\pa_\a Y^k \pa_\a Y^i 
\pa_i Y^l \e_{kl}\over 1- \det \pa_i Y^j}\nn
\pa_i Y^i&=&0\ .
\eea
The first equation comes from noting that $\Gamma_{\a\b}=\gamma_{\a\b}+\dots$
and ensuring that the $\gamma_{\a\b}\e$ terms vanish. 
The next three equations follow from the vanishing of the terms
$\gamma_{\a i}\e$, $\gamma_{12}\e$ and $\e$, respectively.
Using these it is straightforward to show that the $\gamma_\a \rho_i\e$ and
$\gamma_2 \rho_1\e$ terms also vanish. These equations also imply
that 
\bea\label{dipsy}
t_\a &=& -\e_{\a\b\c}\pa_\b Y^1 \pa_\c Y^2\cr
t_i&=&0\ .
\eea
{}From the functional form of $g^{ab}$ we then conclude that 
$t^a=g^{ab} t_b = t_a$. These facts then ensure the vanishing of
the $\gamma_\a\e$, $\gamma_i\e$ and $\rho_i\e$ terms.

To match the two sides of (\ref{bound}) we are thus left to show
that
\be\label{rhs}
{\sqrt{{\det}(g+\tilde H)}}=1-\det \pa_i Y^j +\pa_\a Y^i \pa_\a Y^i +
{\tilde{\cal H}}^{12}(\pa_1 Y^2-\pa_2 Y^1)\ .
\ee              
We first note that the determinant can be rewritten
\be
\det(g_{ab}+\tilde H_{ab})=g+{1\over 2}\tilde{\cal H}^{ab} \tilde{\cal H}_{ab}
+g^{ab}t_a t_b\ .
\ee
and using \p{dipsy} we have
\be 
g^{ab}t_a t_b =(\nabla Y^1\times \nabla Y^2)^2 \ ,
\ee
where $\nabla =(\pa_3,\pa_4,\pa_5)$. We can then 
use Mathematica to show using \p{bpsone} that (\ref{rhs}) 
is indeed an identity. The closure of $H$ imposes one additional
equation:
\be\label{closure}
\pa_\a \pa_\a Y^i =\e_{ij} \pa_j {\tilde{\cal H}}^{12}
\ee

In summary, we have shown that \p{bpsone}, \p{closure} are
the relevant BPS equations corresponding to the configuration
of intersecting branes in \p{second}, 
and that the solutions will preserve 1/8 of the spacetime supersymmetry. 
These BPS equations were first presented in \cite{glwtwo,lwmon}
using the covariant formalism of the fivebrane \cite{hsw}. The
preservation of supersymmetry was checked up to terms quadratic in
the spatial derivatives $\pa_\a$. By contrast here we have checked the
supersymmetry in the Hamiltonian formalism exactly.

The form of $\tilde{\cal H}^{\a1}$ in the BPS equations is somewhat 
analogous to that of a membrane in
the $\{0,1,6\}$ directions and $\tilde {\cal H}^{\a 2}$ that of a membrane in
the $\{0,2,7\}$ directions. More specifically, if we set $Y^2=0$, say, 
then we precisely obtain the BPS equations for a single self-dual string in
the $1$ direction.  Alternatively, if one sets $\pa_\a=0$ and imposes
$\pa_1 Y^2=\pa_2 Y^1$ we get $H=0$ and combined with the last 
equation in \p{bpsone} we are left with the Cauchy-Riemann
equations for $Y^1, Y^2$ as a function of $\s^1,\s^2$,
corresponding to the two intersecting fivebranes.
We thus conclude that the equations generalise the Cauchy-Riemann equations
to include a non-zero $H$ and thus are relevant for describing membranes
ending on complex curves in four dimensions, or equivalently,
a two dimensional special Lagrangian submanifold.
As the equations include the back-reaction of the
membrane on the fivebrane geometry, the fivebrane world-volume
will in general no longer be a complex curve in the $\{1,2,6,7\}$ directions
because only half of the Cauchy-Riemann equations are imposed.
Note that one can impose $\pa_1 Y^2=\pa_2 Y^1$ and still have non-zero $H$
if $\pa_\a\ne0$. These equations are somewhat simpler, but it was shown
in \cite{lwmon} that they do not admit finite energy solutions,
and hence only provide an asymptotic description of the dyons in N=2 SYM
theory. Finally, if one imposes $\pa_i=0$ we obtain the BPS
equations for the delocalised intersecting self-dual strings 
discussed in \cite{glw} and \cite{gkmtz}.

We conclude this section with a comment about the 
energy of solutions solving \p{bpsone}.
It was shown in \cite{ggt} 
that the energy density squared of static fivebrane configurations
is given by
\be\label{energy}
{\cal E}^2 = \det (g_{ab} + \tilde H_{ab}) +t_a t_b m^{ab}\ ,
\ee       
with $m^{ab}=\delta^{ab}-g^{ab}$.
{}For the configurations we have been considering we have
$t_a t_b m^{ab}=0$ and hence 
\bea
{\cal E}&=&{\sqrt{\det(g_{ab} + \tilde H_{ab})}} \nn
&=&{\det g -(\nabla Y^1 \times \nabla Y^2)^2\over 1-\det\pa_i Y^j}\ .
\eea
This last expression differs from that given in \cite{blw} but
agrees with it up to terms quadratic in the spatial derivatives
which was the approximation employed there. Note that if we impose
the other Cauchy-Riemann equation on $Y^1$ and $Y^2$, $\pa_1 Y^2 =\pa_2 Y^1$, 
we obtain
\be
{\cal E}=1+\pa_j Y^i \pa_j Y^i + \pa_\a Y^i \pa_\a Y^i
\ee
as in \cite{blw}.

\section{Three Intersecting Membranes}
We now consider adding another membrane to the array
\p{second} to obtain
\be
\matrix{
M5:&1&2&3&4&5& & & & & \cr
M2:&1& & & & &6& & & & \cr
M2:& &2& & & & &7& & & \cr
M2:& & &3& & & & &8& & \cr}.
\label{blip}
\ee
To preserve supersymmetry we must now impose 
$\gamma^{038}\epsilon=\e$ in addition to \p{gamproj}.
This means that the configuration preserves 1/16 of the
spacetime supersymmetry.
If we denote
$(\gamma^6,\gamma^7,\gamma^8)\equiv (\rho^1,\rho^2,\rho^3)$
then we have 
\bea
\gamma^i\rho^i\e &=& \gamma^1\rho^1\e =
\gamma_{12345}\e\qquad {{\rm no\, sum\, on\,} i}\nn
\gamma^i\rho^j\e &=& -\gamma^j\rho^i\e \qquad i\ne j\ ,
\eea 
where in this section $i,j,\dots=1,2,3$. 
As in the previous section these projections 
imply that we can add several other fivebranes for free. In fact one
finds we can add the fivebranes corresponding to three dimensional
special Lagrangian submanifolds (see section 2.2 of \cite{glw}). 

To realise the intersection of such membranes and fivebranes 
on the first fivebrane we let three
scalars be active: $(X^6,X^7,X^8)\equiv(Y^1,Y^2,Y^3)$ with $Y^i=Y^i(\sigma^j,
\sigma^\a)$, where in this section $\a,\b,\dots =4,5$. 
As before we substitute this into \p{bound} and impose the
projections to determine $\tilde{\cal H}$.
The vanishing of the terms 
with $\e$, $\gamma_{\alpha\b}\e$, $\gamma_{\a i}\e$ and $\gamma_{ij}\e$
imply, respectively, that  
\bea\label{tinky}
\pa_iY^i&=&\det(\pa_i Y^j)\nn
\tilde{\cal H}^{\a\b}&=0&\nn
\tilde{\cal H}^{\a i}&=&M_{ij}\pa_\a Y^j\nn
\tilde{\cal H}^{ij}&=&\e^{ijk}A^k\nn
A^l &=&(\pa_i Y^j+ \pa_\a Y^m\pa_\a Y^i\pa_m Y^j)\e_{ijk}M^{-1}_{kl}
\eea
where the matrix $M$ is defined as
\be
M_{ij}=\delta_{ij}-(-1)^{i+j}\det_{i|j}(\pa_kY^l)
%M_{kl}=\delta_{kl} - 
%{1\over 2}\e^{i_ii_2k}\pa_i_1Y^j_1 \pa_i_2 Y^j_2 \e^{j_1j_2k}
\ee
where the subscript on the $\det_{i|j}$ indicates that we remove the $i$th row
and $j$th column of the matrix $\pa_kY^l$ before taking the determinant.
It is then straightforward to show that the 
$\gamma_\a\rho_i\e$ and $\gamma_{[i}\rho_{j]}\e$ terms automatically vanish
using the lemmas
\bea
\tilde{\cal H}^{\a i}\pa_i Y^j &=&\pa_\a Y^i\pa_i Y^j -\pa_\a Y^j (\pa_l Y^l)\nn
M_{ij} \pa_k Y_j&=& \pa_k Y^i - (\pa_l Y^l)\delta_{ik}
\eea 
One can use Mathematica to show that (\ref{tinky}) implies
\bea
t^k&=& -{1\over 2}\e_{\a\b}\pa_\a Y^i\pa_\b Y^j \e_{ijk}\nn
t^\a&=&\e_{\a\b}\pa_\b Y^i\pa_j Y^k\e_{ijk}
\eea
and hence the vanishing of the $\gamma^k\e$ and $\gamma^\a\e$ terms.
By expanding the identity 
\be
\pa_\a Y^{[i}\pa_\b Y^j\pa^k Y^{l]}=0
\ee
one can then show that
\be
t^i\pa_i Y^l +t^\a \pa_\a Y^l=
-{1\over 2}\e_{\a\b}\pa_\a Y^i\pa_\b Y^j\e_{ijk}\pa_k Y^l
\ee
and hence the $\rho_i\e$ terms vanish.
To complete the proof of supersymmetry one is then left to prove
that
\be
{\sqrt{{\rm det}(g+\tilde H)}}=1-{1\over 2}(\pa_i Y^i)^2+{1\over 2}\pa_i Y^j
\pa_j Y^i+ \tilde{\cal H}^{\a i}\pa_\a Y^i + \tilde{\cal H}^{ij}\pa_iY^j
\ee
This can again be established using Mathematica.
The closure of the three form $H$ then implies that
\be
\pa_\a {\tilde {\cal H}}^{\a j} +\pa_i {\tilde {\cal H}}^{ij}=0 \ .
\ee
The other equation, $\pa_a {\tilde {\cal H}}^{a \b}=0$,
is automatically satisfied by virtue of \p{tinky}.

Let us discuss some special cases of these equations. Firstly we
note that if we impose
\be\label{po}
\pa_i Y^j=\pa_j Y^i\ ,\qquad \pa_\a Y^i=0
\ee
then we have $\tilde {\cal H} =0$ and the equations are simply 
those of a three dimensional special Lagrangian submanifold \cite{hl}. 
The more general equations can thus be seen to include cases
relevant for describing membranes ending on 
such submanifolds. Note that we can just impose 
the first equation in \p{po} leading
to the equations for a three dimensional special Lagrangian manifold
with non-zero three-form. By analogy with the case of two intersecting
membranes one is lead to suspect that this simplification will not
admit finite energy solutions.
 
We next consider the case where $Y^i$ is a function of $\sigma^\a$
only. In this case the self-dual strings are delocalised in the
$\sigma^i$ directions. In this case we find
\bea
{\tilde {\cal H}}^{\a i}&=&\pa_\a Y^i\nn
{\tilde{\cal H}}^{ij}&=&{\tilde {\cal H}}^{\a\b}=0
\eea
and the closure of $H$ then implies that the $Y^i$ are harmonic
functions of two variables. The solutions of these delocalised membranes
then have a logarithmic behaviour. Note that the energy of these solutions
is 
\be
{\cal E}=1+\pa_\a Y^i \pa_\a Y^i\ .
\ee

{}Finally, if we set $Y^3=0$ then the equations become exactly those
of the last section corresponding to two intersecting self-dual
strings.

\section{Four and Five Intersecting Membranes}
The projections imposed on the spinor parameters for the case of four
intersecting membranes with a fivebrane automatically imply the projections 
for a fifth membrane so we shall analyse these cases together. 
The general configuration can be written
\be
\matrix{
M5:&1&2&3&4&5& & & & & \cr
M2:&1& & & &&6&& & & \cr
M2:& &2& & & &&7& & & \cr
M2:& & &3& & &&&8& & \cr
M2:& & &&4& &&&&9& \cr
M2:& & && &5&&&& &\sharp \cr}.
\label{bloop}
\ee
where $\sharp$ represents the number 10.
In addition to \p{gamproj} we must impose 
$\gamma^{038}\e=\e$, $\gamma^{049}\e=\e$ $\gamma^{05\sharp}\e=\e$.
We then have:
\bea
\gamma^a\rho^a\e &=& \gamma^1\rho^1\e =\gamma_{12345}\e\qquad 
{{\rm no\, sum\, on\,} a}\nn
\gamma^a\rho^b\e &=& -\gamma^b\rho^a\e \qquad a\ne b
\eea 
where $(\rho^1,\dots,\rho^5)$$\equiv$$(\gamma^6,\dots,\gamma^\sharp)$
and $a,b,\dots=1,\dots,5$. These projections are those for five dimensional
special Lagrangian submanifolds; the corresponding fivebranes that
one can add to \p{bloop} for free can be found in section 2.4 of \cite{glw}.  
{}For this case all five scalar fields
$(X^6,\dots,X^\sharp)$$\equiv$$(Y^1,\dots,Y^5)$ are excited.
Again we substitute into \p{bound} and demand equality.
The right hand side of \p{bound} becomes
\bea
&&(1-{1\over 2}(\pa_a Y^a)^2+{1\over 2} \pa_a Y^b\pa_b Y^a +\det_{a|a} \pa Y
+{\tilde{\cal H}}^{ab} \pa_a Y^b)\nn
&+&
(\pa_a Y^a -{1\over 2}\Sigma_{a\ne b}\det_{ab|ab} \pa Y + \det \pa Y)
\e^\dagger\gamma^0\e+
(t^e+{1\over 2} \pa_a Y^c \pa_b Y^d \e^{abcde})\e^\dagger\gamma^{0e}\e\nn
&+&(\pa_{[c} Y_{d]} +3 \pa_{[a} Y^c 
\pa _b Y ^a \pa_{d]} Y^b
-{1\over 2}{\tilde{\cal H}}^{cd}+{1\over 2} {\tilde{\cal H}}^{ab}
\pa_a Y^{[c} \pa_ b Y^{d]} )\e^\dagger\gamma^{0cd}\e\nn
&+&({\tilde{\cal H}}^{a[b}\pa_a Y^{c]} - \pa_a Y^a \pa_{[b} Y_{c]} 
+\pa_{[b} Y^a \pa_{|a|} Y_{c]} +(-1)^{b+c}
\det_{b|c}\pa Y)\e^\dagger\gamma^0\gamma_{[b}\rho_{c]}\e\nn
&+&
(t^a\pa_a Y^e+{1\over 2} \e_{b_1b_2b_3b_4b_5}\pa_{b_4} Y^{b_1} 
\pa_{b_5} Y^{b_2} \pa_e Y^{b_3})\e^\dagger\gamma^0\rho_e\e
\eea
where $Y_i\equiv Y^i$.

The vanishing of the $\e^\dagger\gamma^0\e$ terms implies the BPS equation
\be\label{bpsfive}
\pa_a Y^a -{1\over 2}\Sigma_{a\ne b}\det_{ab|ab} \pa Y + \det \pa Y=0
\ee
We next note that the vanishing of the $\e^\dagger\gamma^{0ab}\e$ 
and $\e^\dagger\gamma^0\gamma_{[b}\rho_{c]}\e$
terms imply that 
\bea
{\tilde{\cal H}}^{cd}-{\tilde{\cal H}}^{ab}
\pa_a Y^{c} \pa_ b Y^{d} -2\pa_{[c} Y_{d]} -
3 \pa_{[a} Y^c \pa _b Y ^a \pa_{d]} Y^b
+3 \pa_{[a} Y^d \pa _b Y ^a \pa_{c]} Y^b
&=&0\nn
{\tilde{\cal H}}^{a[b}\pa_a Y^{c]} - \pa_a Y^a \pa_{[b} Y_{c]} 
+\pa_{[b} Y^a \pa_{|a|} Y_{c]} +Z_{bc}&=&0
\eea
where
\be
Z_{bc}={1\over 2} (-1)^{b+c}(\det_{b|c} \pa Y - \det_{c|b} \pa Y)
\ee
If we multiply the second by $\pa_b Y^d$ and substitute into
the first equation we find after some algebra
\be\label{bpsfivetwo}
{\tilde{\cal H}}^{ab}=T_{ac} N^{-1}_{cb}
\ee
where
\bea
N_{ab}&=&\delta_{ab} +\pa_a Y^c\pa_c Y^d\nn
T_{ab}&=&(\pa_a Y^b-\pa_b Y^a)(1-{1\over 2} (\pa_c Y^c)^2 +{1\over 2}
\pa_c Y^d \pa_d Y^c)\nn
&+&(\pa_d Y^c \pa_c Y^a-(\pa_c Y^c)\pa_d Y^a)(\pa_b Y^d -\pa_d Y^b)
-2 Z_{ca} \pa_c Y^b  
\eea
We then need 
\be\label{noonoo}
t^e={1\over 2}\e^{abcde}\pa_aY_b\pa_cY_d
\ee
for the $\e^\dagger\gamma^{0a}\e$ terms to vanish. 
Using the identity
\be
\e^{abcde}\pa_aY_b\pa_cY_d\pa_eY_f=
\e^{abcde}\pa_aY_b\pa_cY_d\pa_fY_e\ ,
\ee
which comes from the identity $\pa_{[a}Y_b\pa_cY_d\pa_eY_{f]}=0$,
we conclude from \p{noonoo} that the
$\e^\dagger\gamma^0\rho_a\e$ terms also vanish. 

We have thus shown that 
solutions to the BPS equations \p{bpsfive} and \p{bpsfivetwo}
together with the closure of $H$, $\pa_a{\tilde{\cal H}}^{ab}=0$,
preserve 1/32 of the supersymmetry provided that
\be
{\sqrt{\det(g+{\tilde H})}}=
1-{1\over 2}(\pa_a Y^a)^2+{1\over 2} \pa_a Y^b\pa_b Y^a +\det_{a|a} \pa Y
+{\tilde{\cal H}}^{ab} \pa_a Y^b
\ee
We have not quite been able to show this with Mathematica using the 
machines available. However, we have made some highly non-trivial checks
and we strongly expect that it is in fact true.
 
Note that if $\pa_aY^b=\pa_b Y^a$ we have ${\tilde{\cal H}}^{ab}=0$ and
the BPS equations truncate to those of a five dimensional special
Lagrangian submanifold \cite{hl}.
We note also that setting $Y^5=0$ will give rise to the BPS equations
for four intersecting membranes associated with a four dimensional
special Lagrangian submanifold,
while if in addition we set $Y^4=0$  we recover the equations obtained
in the last section.
Thus, these BPS equations are the most general of the type
considered in this paper.

\section{Conclusions}
We have constructed BPS equations corresponding to membranes ending on
fivebranes and revealed how they generalise those of special
Lagrangian submanifolds.
The only new solutions that were presented here were those corresponding
to three delocalised intersecting membranes. It would be 
very interesting to construct less trivial solutions, 
but the analysis presented in \cite{lwmon}
for two intersecting membranes indicates that this could be very difficult.

If we dimensionally reduce the general configuration \p{bloop}
on one of the fivebrane directions then we obtain a D-4-brane intersecting
a fundamental string with four D-2-branes. By reducing the BPS
equations that we have derived one will obtain BPS equations for
the abelian Dirac-Born-Infeld theory relevant for these configurations. 
By additional T- and S-duality transformations, one can in principle
obtain BPS equations for many different configurations of intersecting
branes. {}Following \cite{ggt} it was shown in \cite{brecher}
that simpler BPS equations
for the abelian Dirac-Born-Infeld theory have a natural analogue for the
non-abelian theory. Presumably this will also be true for these more
complicated cases.

\vskip 1cm
\noindent{\bf Acknowledgements:} I would like to thank {}Fay Dowker, 
Jose Figueora-O'Farrill and Bill Spence for 
discussions and the EPSRC for partial support. 
 
%%%%%%%%%%%%%%%%%%%%%%%%%%%%%%%%

\end{document}